\documentclass[prl,floatfix,reprint,footinbib,twocolumn,preprintnumbers,amsmath,amssymb,showpacs,superscriptaddress,longbibliography]{revtex4-1} 
\usepackage[T1]{fontenc}
\usepackage[english]{babel}
\usepackage[latin9]{inputenc}
\usepackage{amstext}
\usepackage{graphicx}
\usepackage{esint}
\usepackage{float}
\usepackage{placeins}

\newcommand{\be}{\begin{equation}}
\newcommand{\ee}{\end{equation}}
\newcommand{\ii}{ {\rm i} }
\newcommand{\dd}{ {\rm d} }

\makeatletter
\usepackage[dvipsnames]{xcolor}
\usepackage[unicode=true,
 bookmarks=false,
 breaklinks=false,pdfborder={0 0 1},
 colorlinks=false]{hyperref}
\hypersetup{colorlinks,citecolor=Green,linkcolor=blue,urlcolor=cyan}

\makeatother

\begin{document}

\title{Correlation Functions of the Quantum Sine-Gordon Model in and out of Equilibrium}

\author{I. Kukuljan}
\affiliation{University of Ljubljana, Faculty of Mathematics and Physics, Jadranska ulica 19, SI-1000 Ljubljana, Slovenia}

\author{S. Sotiriadis}
\affiliation{University of Ljubljana, Faculty of Mathematics and Physics, Jadranska ulica 19, SI-1000 Ljubljana, Slovenia}

\author{G. Takacs}
\affiliation{BME ``Momentum'' Statistical Field Theory Research Group, H-1117 Budapest, Budafoki \'{u}t 8, Hungary}
\affiliation{BME Department of Theoretical Physics, H-1117 Budapest, Budafoki \'{u}t 8, Hungary}

\date{\today}

\begin{abstract}
Complete information on the equilibrium behaviour and dynamics of a quantum field theory (QFT) is provided by multipoint correlation functions. However, their theoretical calculation is a challenging problem, even for exactly solvable models. This has recently become an experimentally relevant problem, due to progress in cold-atom experiments simulating QFT models and directly measuring higher order correlations. Here we compute correlation functions of the quantum sine-Gordon model, a prototype integrable model of central interest from both theoretical and experimental points of view. Building upon the so-called Truncated Conformal Space Approach, we numerically construct higher order correlations in a system of finite size in various physical states of experimental relevance, both in and out of equilibrium. We measure deviations from Gaussianity due to the presence of interaction and analyse their dependence on temperature, explaining the experimentally observed crossover between Gaussian and non-Gaussian regimes. We find that correlations of excited states are markedly different from the thermal case, which can be explained by the integrability of the system. We also study dynamics after a quench, observing the effects of the interaction on the time evolution of correlation functions, their spatial dependence, and their non-Gaussianity as measured by the kurtosis.\end{abstract}


\maketitle

\emph{Introduction. - }
Correlation functions provide a complete and practical description
of a physical system. Any observable can be expressed directly in their terms 
and they contain all information about the spectrum of quasiparticles and their collisions \cite{Weinberg}.
In particular, knowledge of higher order correlation functions is necessary
in order to distinguish the ground or thermal states of an interacting from those
of a noninteracting system: while such states are Gaussian for noninteracting
systems, that is, their cumulants (also known as connected correlation functions) of order higher than two vanish,
those of interacting systems are generally 
non-Gaussian.

Recent developments in atom interferometry of ultra-cold atom experiments
have made possible the measurement of correlations of any
order, in both spatial and temporal resolution \cite{exp-sG,exp-GGE}.
Ultra-cold atoms used in these experiments can be confined in elongated
potential traps, so that they form essentially one-dimensional (1D)
quantum gases \cite{1d-phys-rev}. Such gases are gapless systems,
described in terms of their density and phase fields by means of the
\emph{Luttinger liquid} theory \cite{Haldane,Luttinger-liquids}. By splitting the
trap in the transverse direction to form two parallel nearby traps
however, the tunneling between the two traps gives rise to \emph{Josephson
junction} physics \cite{exp:Josephson}. This induces an effective
self-interaction on the phase difference field between the two condensates, such
that its low energy physics is described by the \emph{sine-Gordon} model (SGM) 
\cite{Gritsev2006,Gritsev2007a,Gritsev2007b}, a prominent
example of a strongly correlated quantum field theory (QFT). However, it is still unclear to what
extent this description is valid in out-of-equilibrium settings \cite{new-exp},
like after a \emph{quantum quench}, 
{i.e. an abrupt change of some parameter of the system: 
in such a case, excitations of arbitrarily high energy are typically created.}

The SGM is one of the most studied physical models,
as it describes 2D classical (XY model) \cite{Amit}
and 1D quantum systems (e.g., spin chains). 
It exhibits rich physics 
such as \emph{solitons} and anti-solitons, as well as their bound
states, so-called \emph{breathers}, and it has a topological \emph{Berezinskii-Kosterlitz-Thouless}
phase transition. Both the classical and the quantum versions are integrable
\cite{Faddeev-Korepin_1,Faddeev-Korepin_2}, i.e. 
possess an infinite number of local conserved
quantities allowing exact solution. However, the analytical
calculation of correlation functions is a highly nontrivial
task and {there are only few results available} for higher order correlations:  
{either in special regimes and asymptotic limits (e.g., in the gapless phase or at short or large distances in the gapped phase \cite{Giamarchi,Essler-Konik})} 
or in the classical {thermal} case \cite{exp-sG}, 
on which a comparison with the experiment was based. 
Therefore, the development of numerical methods {is important}, 
both to arrive at theoretical predictions
and for comparison with experimental data in order to test
the validity of the {quantum} SGM description 
{and quantify the relative importance of quantum effects}.

In this Letter, we present an application of the so-called ``\emph{Truncated
Conformal Space Approach}'' (TCSA) \cite{Yurov-Zamo} for the calculation
of SGM correlation functions. The TCSA uses
the \emph{renormalisation group} (RG) \emph{fixed point} of the system
under consideration, as a reference basis for an efficient numerical diagonalisation of the Hamiltonian
\cite{TCSA-3cIsing,TCSA-Smatrix2,TCSA-sG1}.  
It works well in many cases where perturbation theory fails
and is suitable for continuous models in one and even higher
dimensions \cite{TCSA-hd}, unlike DMRG methods that work
principally for lattice 1D models.

While this approach has been used extensively for the study of the
{quantum} SGM \cite{TCSA-sG1,TCSA-sG2,TCSA-sG3}, calculation
of correlation functions has not been achieved until now. 
In the present Letter, we perform this for systems of finite size, as in the experiments.
We calculate two-point (2-p) and four-point (4-p) correlations of the bosonic field 
in coordinate space for different values of the interaction and system
size and in a variety of different settings: ground
states, thermal equilibrium, and excited states, 
as well as quench dynamics. We focus, in particular, on measures
of the non-Gaussianity induced by the interaction. 
In contrast to other approximations for equilibrium \cite{Sachdev, Schweigler1, Schweigler2} 
or quench dynamics \cite{Bertini-Essler, Foini2015, Foini2017a, Foini2017b, DeNardis2017, Kormos1, Kormos2},
our approach us allows to compute multipoint correlations and 
study the full quantum many-body dynamics.

\emph{The sine-Gordon model. -- }
The SGM is a model of a relativistic
interacting bosonic field $\phi$ with Hamiltonian
\[
H_\textrm{SGM}=\int\left(\frac{1}{2}(\partial_{t}\phi)^{2}+\frac{1}{2}(\partial_{x}\phi)^{2}-\frac{m^2}{\beta^2}\cos\beta\phi\right)dx
\]
For $\frac{\beta^{2}}{8\pi} \equiv \Delta >1$ it is gapless, as the cosine potential term 
is an irrelevant perturbation of the free massless
boson Hamiltonian $H_\textrm{FB}=\frac{1}{2}\int(\partial_{\mu}\phi)^{2}dx$
and the field fluctuations are essentially free. Instead, for $\Delta <1$,
the interaction is relevant, the field is locked in one of the cosine minima, and the system 
becomes gapped. The spectrum consists of soliton and antisoliton excitations of mass $M$, 
as well as breathers, whose number is determined by the interaction
parameter $\beta$: the smaller the value of $\beta$, the more breather
modes are present (see the Supplementary Material \cite{SM}). Based on integrability and relativistic invariance \cite{Zamo-Zamo},
the exact particle masses \cite{DHN,Faddeev-Korepin_1},
scattering amplitudes \cite{Zamo77}, vacuum expectation values
\cite{Lukyanov-Zamo1}, and matrix elements of local observables (\emph{form
factors}) \cite{Smirnov} are known. Despite its exceptional solvability properties, however, 
only limited information is available about its correlation
functions. Integrability allows us to compute single point expectation values in finite size and thermal 
\cite{LeClair-Mussardo, Pozsgay2011, Buccheri} or out-of-equilibrium systems \cite{Fioretto, Pozsgay2011, Bertini-Essler,Cubero-Schuricht}
and the infinite volume ground state 2-p function \cite{Essler-Konik}.
However, no {exact} results are available for multipoint observables in thermal or out-of-equilibrium contexts 
with full QFT dynamics.

\emph{TCSA. -- } 
The TCSA, introduced in \cite{Yurov-Zamo} and later applied to the
SGM \cite{TCSA-sG1,TCSA-sG2,TCSA-sG3}, is based on
the idea of using the eigenstate basis of a known reference Hamiltonian
$H_{0}$, truncated up to a specified maximum energy cutoff, 
to construct the ground and excited states of a different Hamiltonian
$H_{0}+V$. The main idea of TCSA and the reason for its
success is to choose as reference Hamiltonian $H_{0}$ the critical
model associated with the UV fixed point of the RG flow that describes
the model under consideration. \emph{Conformal field theory} (CFT)
\cite{CFT} allows the construction of the eigenstates
and energy spectrum of $H_{0}$, as well as the matrix elements of
$V$, and in the truncated basis, $H_{0}+V$ reduces to a finite-dimensional matrix. 
The TCSA is efficient for
computing the energy spectrum of $H_{0}+V$ whenever the perturbation $V$ is relevant; 
it captures effects beyond perturbation theory and does not depend on integrability.

For the SGM, $H_{0}$ is the free massless
boson Hamiltonian $H_\textrm{FB}$ and the perturbing operator $V \propto \int dx\,\cos\beta\phi$
corresponds to the spatial integral of the sum of two so-called \emph{vertex
operators} $\exp\left(\pm\mathrm{i}\beta\phi\right)$. Here we consider a 
finite system of size $L$ with Dirichlet boundary conditions $\phi(0)=\phi(L)=0$,
which preserve integrability \cite{Ghoshal-Zamo} and induce well-understood changes 
to the energy spectrum \cite{bsG1,TCSA-bsG}.This boundary condition significantly simplifies 
our computations with TCSA, since the truncated Hilbert space consists of a single Fock space
independent of $\beta$ \cite{SM}.
Our numerics are validated by comparing with the known
mass spectrum \cite{DHN,Faddeev-Korepin_1} 
and one-point observables \cite{Lukyanov-Zamo1,Buccheri} (see also the Supplementary Material \cite{SM}).

\emph{Correlations in equilibrium states. - } 
We start our analysis with equilibrium states of the SGM, i.e. the ground state (the minimum-energy eigenstate of $H_{SGM}$ for chosen parameter values), 
thermal states (defined through the density matrix $\rho \propto e^{-H_{SGM}/T}$ where $T$ the temperature) and excited states (eigenstates of $H_{SGM}$ with energy higher than the ground state). 
We compute equal time correlation functions 
\be
G^{(N)}(x_1,x_2,\ldots,x_N)=\left\langle\phi(x_1)\phi(x_2)\cdots\phi(x_N)\right\rangle, \label{eq:cf}
\ee
as well as their connected part $G_{\text{con}}^{(N)}(x_1,
...,x_N)= \sum_{\pi}\left(|\pi|-1\right)!(-1)^{|\pi|-1}\prod_{B\in\pi} {\left\langle\prod_{i\in B}\phi(x_i)\right\rangle}$, 
where $\pi$ are all partitions of $\{1,2,\ldots,N\}$ into blocks $B$, $i$ are elements of $B$, and $|\pi|$ is the number of blocks in $\pi$ \cite{SM}. 
In the above definitions, the expectation value refers to the state under consideration in each case. 
We focus on the quantities measured in the experiments: the 2-p and 4-p full and connected correlation functions, as well as the \emph{kurtosis} (the ratio of connected over full 4-p correlations integrated over all space {$\mathcal{K}={\int \mathrm{d}V |G_{\text{con}}^{(4)}|}/{\int \mathrm{d}V |G^{(4)}|}$, with $ \int \mathrm{d}V = \prod_{i=1}^4 \int_0^L\dd x_i$}) 
{which measures how 4-p correlations deviate from those of a Gaussian state}. 

\begin{figure}[htbp]
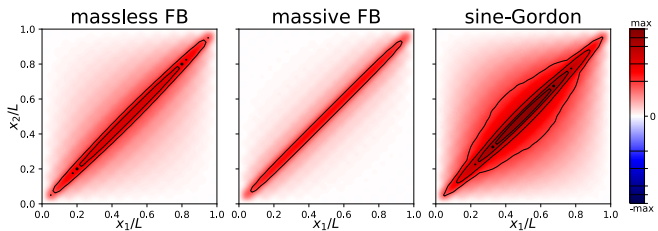

	\centering
	\includegraphics[width=\linewidth]{{{GroundStates}}}
	\caption{Density plots of 2-p correlations $G^{(2)}(x_1,x_2)$ on the ground state of the SGM in a box (\emph{right})  in comparison with those of the massless (\emph{left}) and massive (\emph{middle}) free boson (FB) case [interaction $\Delta=1/18$, system size $L=25$  (in units of $M$), mass of the free case chosen equal to first breather mass of the SGM].}
	\label{fig:GS}
\end{figure}

\begin{figure}[htbp]
	\begin{center}
		\centering
		\includegraphics[width=\linewidth]{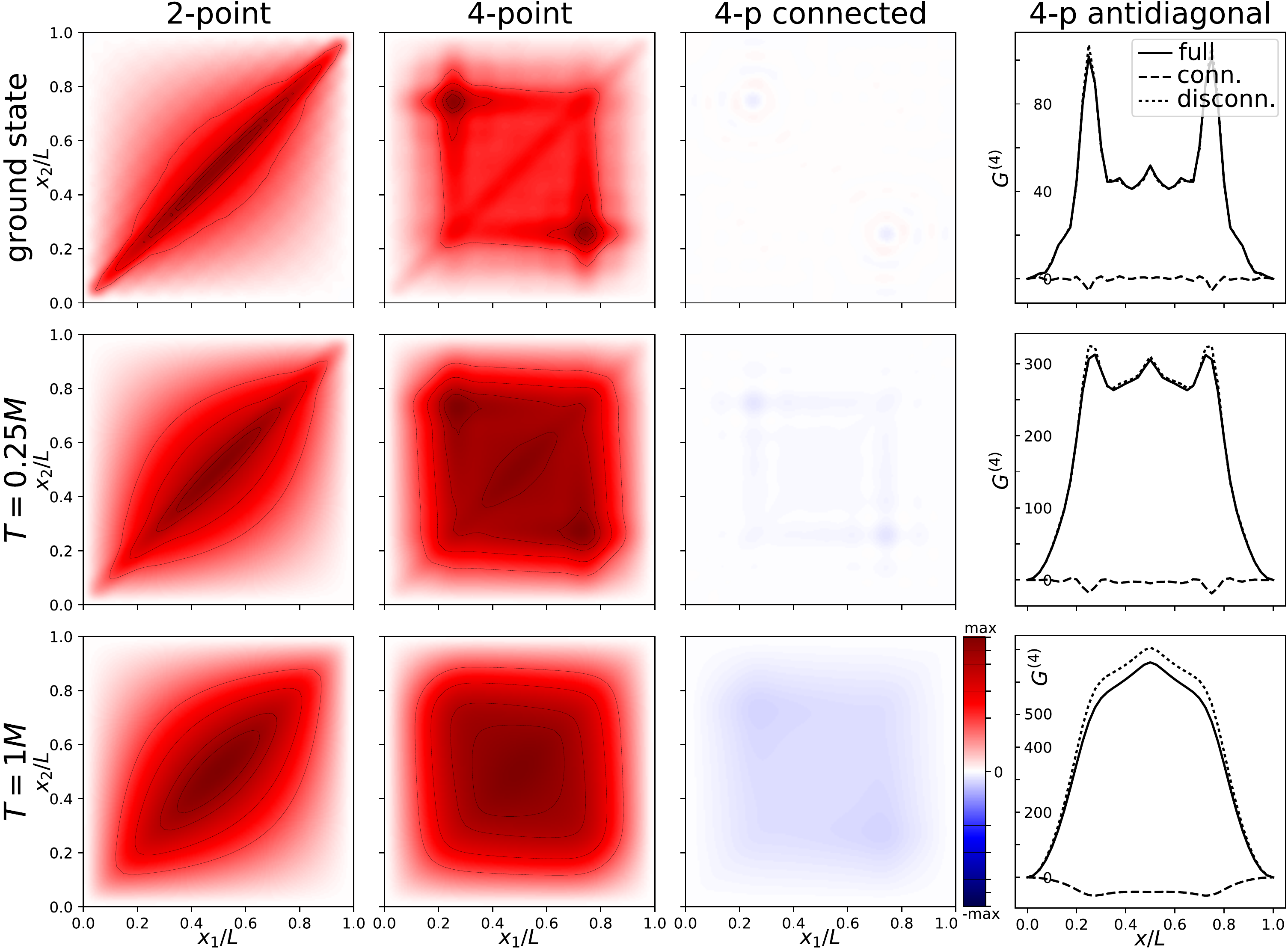}
		\caption{Density plots of 2-p correlations $G^{(2)}(x_1,x_2)$ and 4-p full 
			and connected correlations $G^{(4)}_{\text{(con)}}(x_1,x_2,L/4,3L/4)$ for the ground and two thermal states ($\Delta=1/18$, $L=25$, $T=0,\,0.25,\,1$ in units $M$). {For better comparison, the last column shows plots of the 4-p full (\emph{solid line}), connected (\emph{dashed}) and disconnected (\emph{dotted}) correlations along the antidiagonal section $x_2=L-x_1$.}} 
		\label{fig:eq-st}
	\end{center}
\end{figure}

Figures \ref{fig:GS} and \ref{fig:eq-st} 
show typical plots of 2-p and 4-p functions in ground and thermal states
of the SGM at interaction $\Delta=1/{18}\approx 0.055$, which is 
in the highly attractive regime, similar to the experimentally realised system,
and well inside the window where our numerics
can reliably produce a large part of the excitation spectrum. 

The results can be compared with those of the free massless or free massive case with mass equal to the lowest breather mass (Fig.~\ref{fig:GS}): switching on the interaction results in dramatic growth of ground state correlations that are also longer in range than in the free case. 
At the same time 4-p connected correlations appear, signaling the non-Gaussianity of the state, 
albeit in ground states they are small in comparison with the full 4-p correlations.
Larger deviations from Gaussianity are seen in thermal or excited states, which is explained by the following semiclassical argument.
The ground state energy is close to the bottom of the cosine potential, where it is well approximated by a parabola. In contrast, finite energy density of thermal states allows exploring the nonparabolic shape of the potential \cite{SM}. At high temperatures, correlations are dominated by excitations with energy well above the potential, which are essentially free massless bosonic modes and so non-Gaussianity is suppressed.

This is demonstrated in Fig.~\ref{fig:kurt-temp} showing the kurtosis $\mathcal{K}$ 
as a function of the temperature $T$: its maximum value is observed at temperatures 
comparable to the height of the cosine potential, i.e., $T\sim M$. 
This thermal effect is precisely what gives rise to the experimental observation 
of three different regimes of the SGM \cite{exp-sG}: 
free massless phonons (high $T$), 
coexistence of interacting massive phonons and solitons (intermediate $T$), 
and free massive phonons (low $T$). 
Note that even though $\mathcal{K}$ decreases for increasing $T\gtrsim M$ 
as the free high energy excitations contribute more and more, 
its precise value in the limit $T\to\infty$, which is inaccessible by TCSA, is not 
necessarily zero as the low energy excitations still have a nonzero contribution. 
Notice also how increasing $T$ results in correlations being less concentrated along the diagonal (Fig.~\ref{fig:eq-st}).

While ground and thermal states exhibit a rather simple pattern, 
characterised by decay of correlations with separation distance, 
excited states display much richer structure as shown in Fig.~\ref{fig:exc-st}. 
The strong qualitative difference in correlations between excited and thermal states 
and between excited states even at nearby energies may be seen as
a violation of the \emph{eigenstate thermalisation hypothesis} \cite{Deutsch,Srednicki,Rigol-Srednicki} for the SGM: 
due to its integrability, a typical eigenstate exhibits local characteristics 
dramatically different from those of a thermal state with the same energy density. 

\begin{figure}[htbp]
	\begin{center}
		\centering
		\includegraphics[width=\linewidth]{{{kurtosis}}}
		\caption{Kurtosis (measure of non-Gaussianity) in thermal states of the quantum sine-Gordon model, as a function of the inverse temperature $T^{-1}$ for interactions $\Delta\equiv \beta^2/(8\pi)=1/100$ (red line) and 1/18 (green line). Deviations from Gaussianity increase with $T$ up to a maximum at $T\sim 1$ (in units of soliton mass $M$) and decrease at higher $T$ since the high energy states are essentially free. Our method describes accurately the low and intermediate temperature regime $T\lesssim 1$. Insets illustrate the excitation level relative to
			the height of the cosine potential (see the Supplementary Material \cite{SM}).\label{fig:kurt-temp}}
	\end{center}
\end{figure}

\begin{figure}[!htbp]
	\begin{center}
		\centering
		\includegraphics[width=\linewidth]{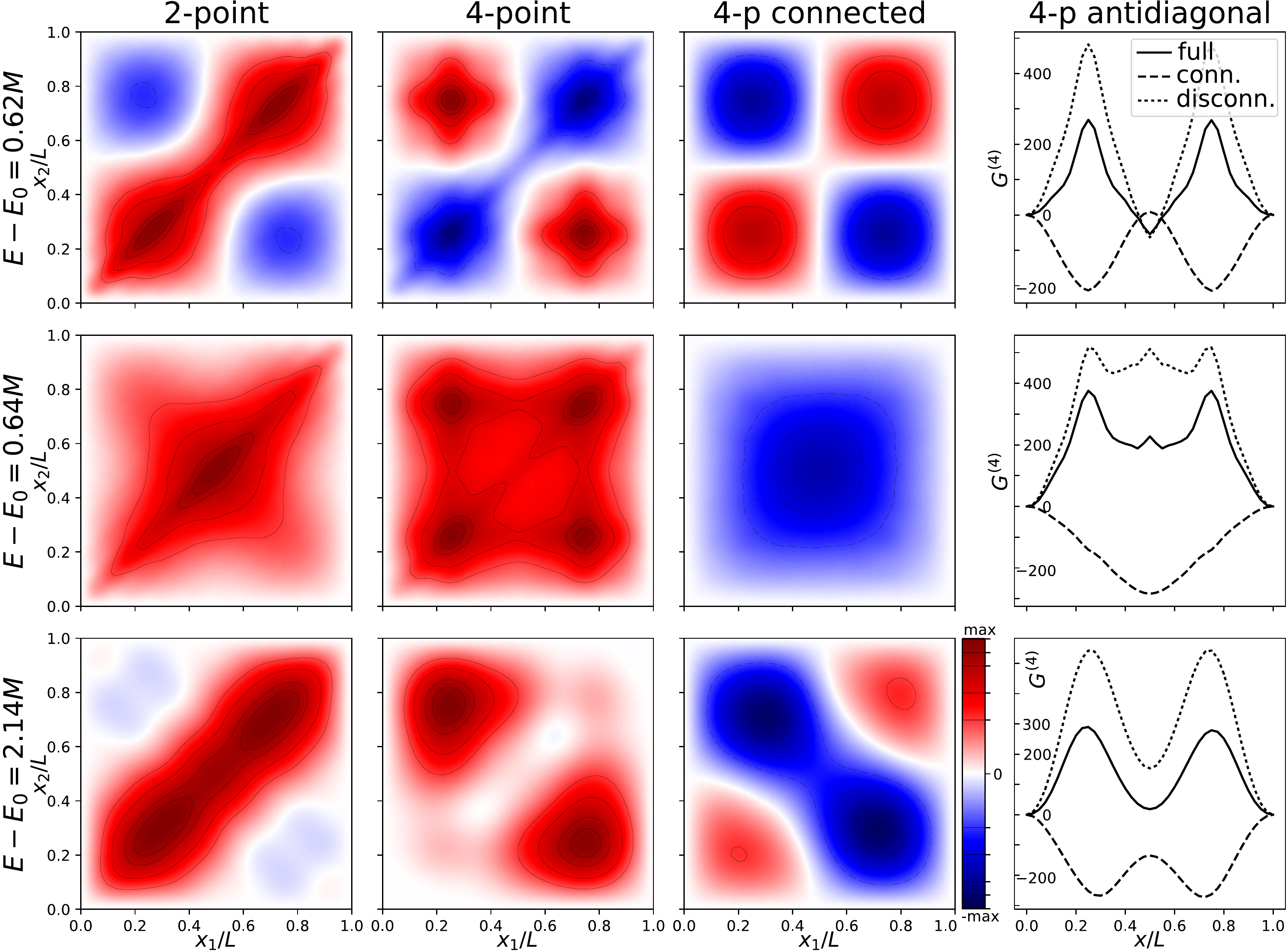}
		\caption{\label{fig:exc-st}Density plots of 2-p, 4-p full and connected correlations for some excited states. Note the strong qualitative differences in the patterns, even at nearby energy levels (\emph{top and middle row}). {Last column: comparative plots of 4-p antidiagonal correlations.}}
	\end{center}
\end{figure}

\begin{figure*}[ht]
	\begin{center}
		\centering
		\includegraphics[width=\textwidth]{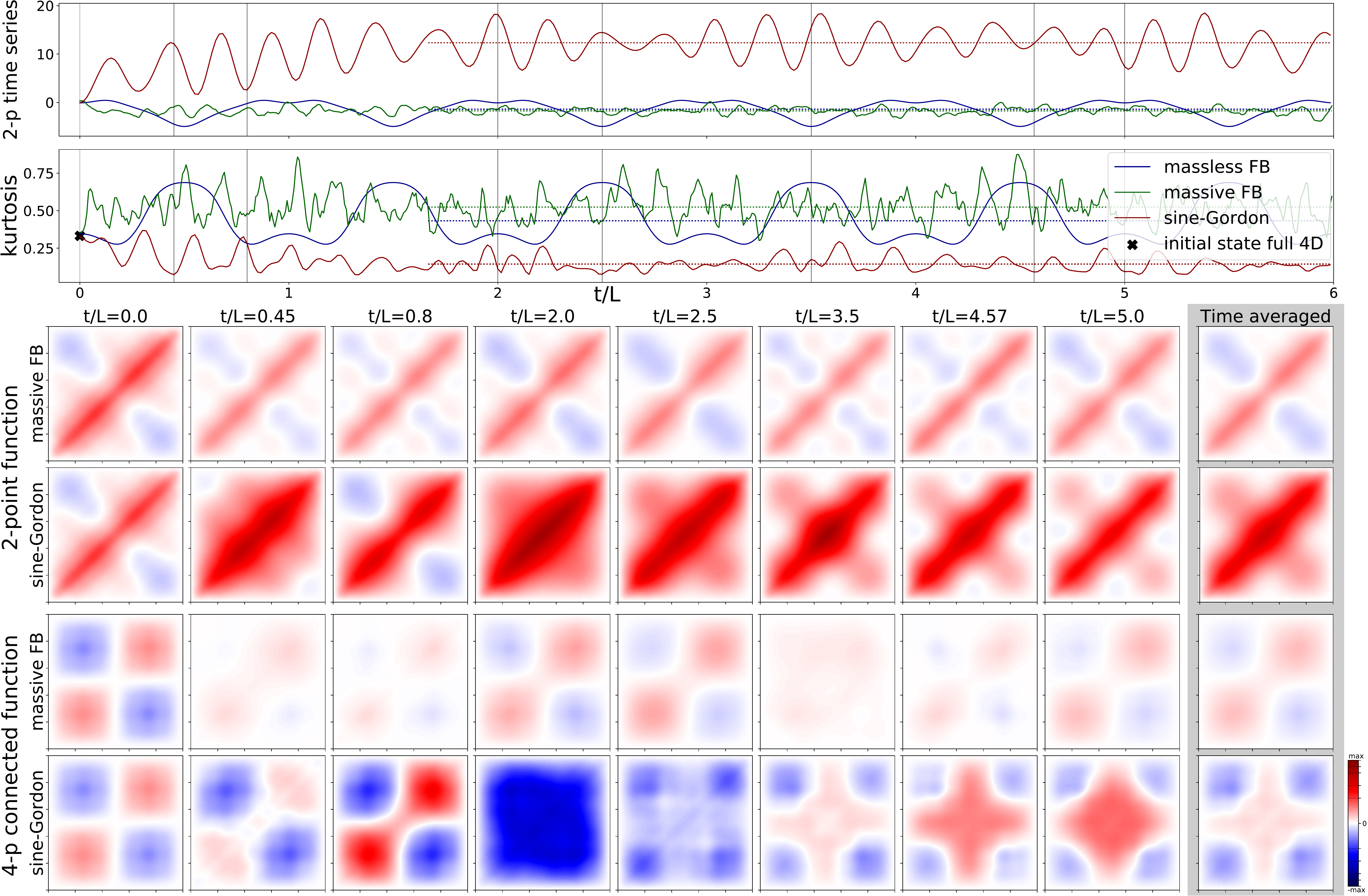}
		\caption{Time evolution of 2-p correlations $G^{(2)}(L/3,2L/3;t)$ (\emph{top row}), kurtosis (\emph{second row}), and spatial density plots of 2-p correlations $G^{(2)}(x_1,x_2;t)$ and 4-p correlations $G^{(4)}_{\text{con}}(x_1,x_2,L/4,3L/4;t)$ at various times $t$ (\emph{bottom four rows}) and time averaged (\emph{last column}) after a quench starting from an excited state of the SGM (prequench interaction $\Delta_0 =1/{18}$, postquench interaction $\Delta =1/{8}$, initial state energy $\sim 0.73 M$ above ground state, corresponding to 50\% of the height of the cosine potential, $L=30/M$). Correlations under massless and massive free boson dynamics (with mass matched to that of the first breather of the SGM) are plotted together for comparison. (Density plot axes correspond to $x_1/L$ and $x_2/L$ as in Figs.~\ref{fig:eq-st} and \ref{fig:exc-st}. For an animation of these data and technical details about the computation of the dynamics of the kurtosis, see 
			the following url: \url{https://arxiv.org/src/1802.08696/anc/animation.mp4}.)} 
		\label{fig:quench}
	\end{center}
\end{figure*}

\emph{Correlations in out-of-equilibrium states. -- } 
We now consider the dynamics
after a quantum quench, i.e., the time evolution of correlations when the system is initially prepared in a state $|\Psi \rangle$ that is an equilibrium 
state of some Hamiltonian $H_0$ and evolved with a different Hamiltonian $H$ for time $t$. In this dynamical case, the correlation functions, denoted as $G^{(N)}(x_1,x_2,\ldots,x_N;t)$,
are still given by (\ref{eq:cf}) but with the expectation value referring to the time evolved state $e^{-\ii H t} |\Psi \rangle$. 
We mainly focus on the case where the initial state is 
an excited state of the SGM for some value of the interaction ($\Delta_0=1/{18}$) and 
the time evolution corresponds to another value ($\Delta=1/{8}$). 
The excited state we used corresponds to half of the height of the cosine potential of the prequench Hamiltonian, that is, far from the bottom and deeply in the nonparabolic regime.
Choosing an excited rather than ground or thermal state as initial state 
results in a higher value of postquench energy density 
and in more interesting dynamics \cite{SM}, which is
beyond the regime of validity of low density or semiclassical approximations.

Figure \ref{fig:quench} shows the time evolution of the 2-p function 
at two fixed points and of the kurtosis,
as well as snapshots of the spatial dependence of 2-p and 4-p connected functions at various 
times and after time averaging. 
For comparison, we present also the time evolution of the same initial state under 
free massless and massive dynamics. 

The dynamics shows strong oscillations in all three cases;
however, despite initiating from the same state, the three types of dynamics 
result in different oscillation frequencies, time averages, and amplitude of fluctuations
about the average. The free massless case is an exception 
in that the dynamics is purely periodic with revival period $t_R=L$ since all energy differences
are integer multiples of $2\pi/L$ due to linear dispersion. 
In the free massive case, exact periodicity is lost as particle velocity depends on the momentum. 
Instead, we observe phase oscillations with frequencies dominated by the particle mass.
SGM dynamics is characterised by multifrequency oscillations \cite{Gritsev2007a}, a general amplification of the initial correlations compared to the free case and a qualitative pattern change (from an initial checkerboard to a cross, in the case of 4-p correlations).

The dynamics of kurtosis reveals another interesting property. 
In contrast to the free cases where it shows strong fluctuations and passes frequently close to its initial value, 
its SGM dynamics is characterised by
a long plateau different from its initial value, with relatively small fluctuations about the time average.
This observation points to the concept of \emph{equilibration on average} \cite{vonNeumann1929,vonNeumann,Gogolin-Eisert} 
that applies to finite systems. Note that spatially integrated local observables like the kurtosis are essentially global measures of correlations.
The difference between free and interacting dynamics of the kurtosis can be attributed to the fact that in the free case the $\phi$ field is related
in a simple linear way to the system's diagonal modes. On the contrary, in the interacting case, its mode (or form factor) expansion is 
intrinsically far more complicated, resulting in efficient dephasing, even for such a global quantity.

By extensive numerical experimentation,
we have checked that the dynamical behaviour presented here is typical 
and robust with respect to the choice of initial state and quench parameters in the regime we study.

\emph{Discussion. -- } 
We have demonstrated that truncated Hamiltonian methods can be efficiently applied to compute multipoint correlation functions in QFT both in and out of equilibrium, allowing also access to non-Gaussianity measures such as the kurtosis.  For the sine-Gordon model we observe that quench dynamics changes the spatial pattern of the connected 4-p correlations substantially, which is in marked contrast with the free case and is a nontrivial effect of interactions. Excited state connected 4-p correlations were also found to be significantly different from the thermal ones, which can be understood on the basis of integrability. 

\begin{acknowledgments}
G.T. is grateful to J. Schmiedmayer for an enlightening discussion about their experiment. The authors also thank A. Tsvelik for useful comments on the manuscript. The work was partially supported by the Advanced Grant of European Research Council (ERC) 694544 -- OMNES, by the Slovenian Research Agency under grants N1-0025, N1-0055 and P1-0044, by the National Research Development and Innovation Office of Hungary within the Quantum Technology National Excellence Program (Project No. 2017-1.2.1-NKP-2017-00001) and under OTKA grant No. SNN118028 and also by the BME-Nanotechnology FIKP grant of EMMI (BME FIKP-NAT).
\end{acknowledgments}



\onecolumngrid

\FloatBarrier

\

\appendix

\begin{center}
\Large{\bf{Supplementary Material}}
\end{center}

\section{The sine-Gordon model spectrum}

\paragraph{Mass spectrum -}

The action of the sine-Gordon model as a perturbed conformal field theory can be written as:
\be
\mathcal{S}_{\text{SGM}}=\int_{-\infty}^{\infty}\dd t\int_{0}^{L}\dd x\, \left[\frac{1}{8\pi}\partial_\nu \varphi \partial^\nu \varphi+\mu : \cos\left(\frac{\beta}{\sqrt{4\pi}}\varphi\right) : \right]\label{eq:SGactionZamolodchikov}
\ee
where the semicolon denotes normal ordering of the free massless boson modes. The relation between the soliton mass $M$ and the coupling parameter $\mu$ is \cite{Zamo1995}:
\be
\mu=\kappa(\xi)M^{2/(\xi+1)},
\ee
where the parameter $\xi$ is defined as: 
\be
\xi=\frac{\beta^2}{8\pi-\beta^2}=\frac{\Delta}{1-\Delta},\label{eq:DefinitionP}
\ee
with $\Delta$ the conformal weight of the vertex operator and the coupling-mass ratio $\kappa(\xi)$ is \cite{Zamo1995}:
\be
\kappa(\xi)=\frac{2}{\pi}\frac{\Gamma\left(\frac{\xi}{\xi+1}\right)}{\Gamma\left(\frac{1}{\xi+1}\right)}\left[\frac{\sqrt{\pi}\Gamma\left(\frac{\xi+1}{2}\right)}{2\Gamma\left(\frac{\xi}{2}\right)}\right]^{2/(\xi+1)}.
\ee
In \eqref{eq:SGactionZamolodchikov} we have used the rescaled field:
\be
\phi=:\frac{1}{\sqrt{4\pi}}\varphi
\ee
and compactified it on a circle of radius $R$:
\be
\varphi\sim\varphi+2\pi R, \qquad R=\frac{ \sqrt{4\pi}}{\beta}=\sqrt{\frac{\xi+1}{2\xi}}=\frac{1}{\sqrt{2\Delta}} \label{eq:Compactif}
\ee
to take into account the periodicity of the cosine potential. 
In the above the length of the system is $L$ and we impose Dirichlet boundary conditions:
\be
\varphi(0)=\varphi(L)=0
\ee

\begin{figure}[htbp]
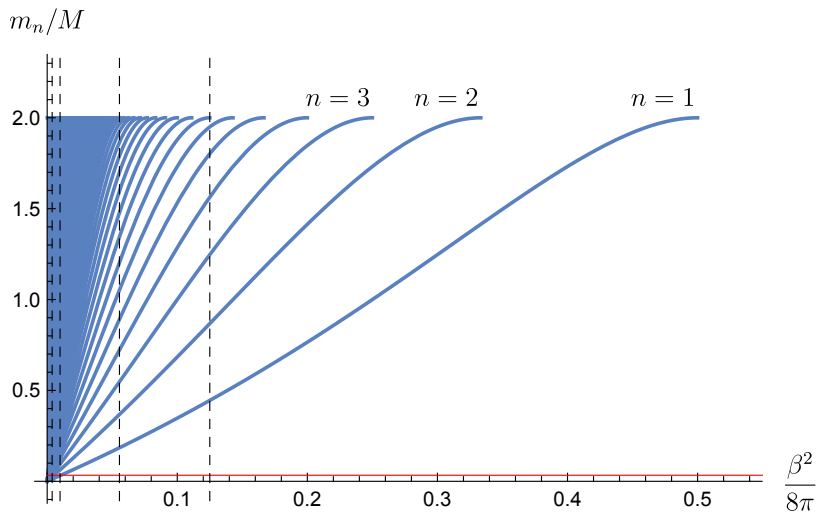

\begin{center}
 \centering
\includegraphics[width=.6\linewidth]{{{sG_masses}}}
\caption{Breather masses $m_n$ in units of soliton mass $M$ as a function of interaction $\beta^2/(8\pi)$. Dashed vertical lines denote the values of interaction used for the numerics: from left to right $\frac1{260},\frac1{100},\frac1{18}$ and $\frac1{8}$. The horizontal line denotes the inverse system size $1/(ML)=1/25$ for comparison.}
\label{fig:sG-masses}
\end{center}
\end{figure}

The SGM particle spectrum consists of solitons, anti-solitons and for $\xi<1$ also breathers, i.e. soliton-antisoliton bound states. For a given $\xi$, there are $n=1,2,...<1/\xi$ breathers with masses:
\be
m_n = 2 M \sin (n \pi \xi/2)
\ee
plotted as a function of the interaction $\Delta=\beta^2/(8\pi)=\xi/(\xi+1)$ in Fig.~\ref{fig:sG-masses}.

\paragraph{Excitation level of states -}

Here we discuss in more detail the argument used in the main text to explain the behaviour of non-Gaussianity of the states investigated. For any state it is possible to express its excitation level compared to the potential using a single dimensionless quantity. The potential term of the SGM Hamiltonian is
\be
-\int dx\frac{m^{2}}{\beta^{2}}\cos\beta\varphi
\ee
and for small $\beta$, the soliton mass is
\be
M=\frac{8m}{\beta^{2}}
\ee
Assuming that the energy of the state relative to the ground state is given by $\chi$
\be
E-E_{0}=\chi M
\ee
while the potential height in finite volume is
\be
\Delta V=\frac{2m^{2}}{\beta^{2}}L
\ee
the relevant ratio is given by
\be
\frac{E-E_{0}}{\Delta V}=\frac{\chi M}{\frac{2m^{2}}{\beta^{2}}L}=\frac{4\chi}{mL}
\label{energyratio}
\ee
where $m$ can be replaced with the first breather mass $m_{1}$ in the small $\beta$ regime.

If this dimensionless quantity is small, the state is lying at the bottom of the potential, where it is effectively parabolic. Thus, the excitations in such a state are free massive phonons and non-Gaussianity is suppressed. This happens in the ground state and low-temperature states. On the contrary, if the dimensionless ratio is higher, of the order of $0.5$, then the state experiences the full cosine potential, the excitations are solitons and breathers and the state can be highly non-Gaussian. This happens at intermediate temperatures and in the low-energy excited states. If the dimensionless ratio is even higher, that is much higher than $1$, the cosine potential becomes insignificant and the system is effectively free - the excitations are free massless bosons and non-Gaussianity is again suppressed. This happens at high temperatures and in highly excited states.

\section{Truncated Conformal Space Approach for the  sine-Gordon correlation functions\label{app:TCSA}}

In this section we explain the adaptation of the Truncated Conformal Space Approach (TCSA) to compute the correlation functions of the sine-Gordon model on a finite interval.  The general idea of the TCSA is to write the theory of interest as the conformal part plus a relevant perturbation. Then, the Hamiltonian and the observables are expressed as matrices in the basis of the conformal part and  a truncation at certain energy is introduced to keep the matrices finite. All the operators are expressed in the standard CFT notation of complex Euclidean spacetime coordinates $z=e^{\frac{\pi}{L}(\tau-\ii x)}$ and $\bar{z}=e^{\frac{\pi}{L}(\tau+\ii x)}$ with $\tau=\ii t$ the imaginary time. However, the complex coordinates are introduced just to aid the computation of the matrix elements of the operators in the CFT basis. The TCSA time propagation in our work is always done in real time $t$ and we use the Schr\"{o}dinger picture in which all operators expressing fields and physical observables are time independent and only the states carry the time evolution. Therefore we can always use the expressions for the operators at $\tau=0$.
	
Let us begin by introducing the vertex operator defined as \cite{CFT}:
\be
V_n(z,\bar{z})=e^{i\frac{n}{R}\varphi(z,\bar{z})}.
\ee
With its help, we can write the sine-Gordon Hamiltonian as:

\begin{eqnarray}
H_{\text{SGM}}
&:=& H_{\text{FB}}- \frac{\kappa(\xi)}{2} \left(\frac{\pi}{ML}\right)^{2\Delta} M^2  \int_{0}^{L}\dd x \left(V_{1}(e^{-\ii\frac{\pi}{L}x},e^{\ii\frac{\pi}{L}x})+V_{-1}(e^{-\ii\frac{\pi}{L}x},e^{\ii\frac{\pi}{L}x})\right)\label{eq:SGhamiltonian}
\end{eqnarray}
where  

\be
H_{\text{FB}}=\frac{1}{8\pi}\int_{0}^{L}\dd x \left[\left(\partial_t \varphi\right)^2+\left(\partial_x \varphi\right)^2\right]
\ee
is the free massless boson Hamiltonian. The soliton mass $M$ plays the role of energy or inverse length unit. The factor $\left(\frac{\pi}{ML}\right)^{2\Delta}$ is the conformal scaling factor associated with the vertex operator when transforming from the strip of width $L$ to the plane geometry  and the corresponding scaling dimension is $\Delta$ \cite{CFT}.


The Hamiltonian \eqref{eq:SGhamiltonian} is already written in the TCSA form, where the free massless boson term represents the exactly solvable (conformal) part and the interaction term represents the relevant perturbation. In the following, we discuss the free massless boson Hilbert space, give the matrix elements of the operators used in the computation and explain how the TCSA simulation is done. 

\paragraph{Free massless boson Hilbert space -} The boson field $\varphi$ satisfying Dirichlet boundary conditions takes the form \cite{CFT}:
\be
\varphi(x)=\varphi_0- \frac{2\pi}{L}R W x+2\sum_{k\neq0}\frac{a_k}{k}\sin(k{\pi}x/{L}).\label{eq:FBfield}
\ee
For other boundary conditions (Neumann and periodic), $\varphi_0$ is an operator, which is divergent in itself and only its exponential is well defined. In the case of Dirichlet boundary conditions, $\varphi_0$ is a number corresponding to the $x=0$ boundary value of the field and in our case $\varphi_0=0$.  The operator $W$ gives the difference of the field at the two ends of the interval; for the case of periodic boundary conditions its values are quantised and give the winding number of the compact scalar field, a.k.a. the topological charge carried by the solitonic excitations. For Dirichlet boundary conditions, different values of $W$ correspond to distinct sectors consisting of a single Fock space. The case of two identical Dirichlet boundaries on the strip corresponds to $W=0$.

We quantize the field using the following commutation relations:
\be
\left[a_k,a_l\right]=k\delta_{k+l}.\label{eq:CommRelGeneral}
\ee
From the vacuum state $\left|0\right>$ that is defined by:
\be
a_k\left|0\right>=0,\quad\text{for all }k>0\label{eq:Annih}
\ee
we can construct descendant states by acting with the creation operators $a_{-k}=a_k^\dagger$, a (nonnegative integer) number of times $r_k\geq 0$:
\be
\left|\vec{r}\right>=\left|r_1,r_2,\ldots,r_k,\ldots\right>:=\frac{1}{N_{\vec{r}}}\prod_{k=1}^{\infty}a_{-k}^{r_k}\left|0\right>,\label{eq:States}
\ee
The normalization is given by: 
\be
N_{\vec{r}}^2=\left<0\right|\prod_{k=1}^{\infty}a_k^{r_k}a_{-k}^{r_k}\left|0\right>=\prod_{k=1}^{\infty}(r_k!k^{r_k}).\label{eq:Normalization}
\ee
These states provide a basis of the $W=0$ Fock space, that is the Hilbert space for our problem.

\paragraph{Matrix elements -}
To perform the TCSA, we have to compute in the free boson Hilbert space the matrix elements of all operators needed in the computation. This is done by preforming the algebra using the commutation relations \eqref{eq:CommRelGeneral}. For a given pair of basis states of the $W=0$ Fock space:
\begin{eqnarray}
\left|\psi\right>&=&\left|\vec{r}\right>,\\
\left|\psi'\right>&=&\left|\vec{r}'\right>,
\end{eqnarray}
let us denote the corresponding matrix element of an operator $O$ by:
\be
O^{\psi',\psi}:=\left<\psi'\right|O\left|\psi\right>.
\ee
For the results presented in this work we need the following operators.

The free massless boson Hamiltonian for Dirichlet boundary conditions is diagonal with matrix elements: 
\be
H_{\text{FB}}^{\psi',\psi}=\frac{\pi}{L}\left(
\sum_{k=1}^{\infty}kr_k-\frac{1}{24}\right)\delta_{\psi',\psi}.\label{eq:FBmatrixElements}
\ee

The Hamiltonian of the massive free boson:
\be
H_{\text{mFB}}=\frac{1}{8\pi}\int_{0}^{L}\dd x \left[\left(\partial_t \varphi\right)^2+\left(\partial_x \varphi\right)^2+m^2\varphi^2\right].
\ee
has the following matrix elements:
\begin{eqnarray}
H_{\text{mFB}}^{\psi',\psi}&=&\frac{\pi}{L}\left\{\delta_{\psi',\psi}\left(\sum_{k=1}^{\infty}\left(1+\frac{m^2L^2}{2\pi^2k^2}\right)kr_k-\frac{1}{24}\right)+\right.\\
&&\left.+\frac{m^2L^2}{4\pi^2}\sum_{k=1}^{\infty}\left(\prod_{n=1\atop n\neq k}^{\infty}\delta_{r'_n,r_n}\right)\frac{1}{k^2}\left(\sqrt{r_k k}\sqrt{(r_k-1) k}\,\delta_{r'_k+2,r_k}
+\sqrt{(r_k+2) k}\sqrt{(r_k+1) k}\,\delta_{r'_k-2,r_k}
\right)\right\}.\nonumber
\end{eqnarray}

The expression for the vertex operator can be written in normal ordered form as \cite{CFT}:
\be
V_n(z,\bar{z})=e^{iq\varphi(z,\bar{z})}=\left|z-\bar{z}\right|^{-q^2}:e^{iq\varphi(z,\bar{z})}:
\ee
where $q\equiv n/R$ with the value of the compactification radius $R$ given in (\ref{eq:Compactif}). 
Its matrix elements are:
\be
V_n^{\psi',\psi}\left(e^{\ii\frac{\pi}{L}x},e^{-\ii\frac{\pi}{L}x}\right)=N_{\vec{r}'}^{-1}N_{\vec{r}}^{-1}\left[2\sin\left(\frac{\pi x}{L}\right)\right]^{-q^2}\prod_{k=1}^{\infty}\left<0\right|a_k^{r'_k}e^{- q \frac{a_{-k}}{k}(z^k-\bar{z}^k)}e^{ q \frac{a_k}{k}(z^{-k}-\bar{z}^{-k})}a_{-k}^{r_k}\left|0\right>,
\ee
with:
\begin{eqnarray}
&\left<0\right|a_k^{r'_k}e^{- q \frac{a_{-k}}{k}(z^k-\bar{z}^k)}e^{ q \frac{a_k}{k}(z^{-k}-\bar{z}^{-k})}a_{-k}^{r_k}\left|0\right>=&\nonumber\\
&=\sum_{j'=0}^{\infty}\sum_{j=0}^{\infty}\frac{1}{j'!j!}\left(\frac{2q}{k}\right)^{j'+j}\left[\frac{\bar{z}^k-z^k}{2}\right]^{j'+j}\left<0\right|a_k^{r'_k}a_{-k}^{j'}a_k^{j}a_{-k}^{r_k}\left|0\right>&
\end{eqnarray}
and:
\be
\left<0\right|a_k^{r'_k}a_{-k}^{j'}a_k^{j}a_{-k}^{r_k}\left|0\right>=k^{j'+j}\left(\begin{array}{c}r'_k\\j'\end{array}\right)\left(\begin{array}{c}r_k\\j\end{array}\right)j'!j!(r_k-j)!k^{r_k-j}\delta_{r'_k-j',r_k-j}\Theta(r_k\geq j).\label{eq:CombinatoricTerm2}
\ee

To get the matrix elements of the spatially integrated vertex operator that appears in the sine-Gordon Hamiltonian (\ref{eq:SGhamiltonian}), the following relation is useful:

\be
\int_{0}^{\pi}\dd u \left[2\sin\left(u\right)\right]^{-q^2}e^{-\ii k u}=\frac{e^{-\ii\frac{\pi}{2}k}\pi}{(1-q^2)B\left(\frac{1}{2}(2-q^2-k),\frac{1}{2}(2-q^2+k)\right)}.
\ee
Here, $B(x,y)=\frac{\Gamma(x)\Gamma(y)}{\Gamma(x+y)}$ is the beta function.

Lastly the matrix elements of the $\varphi$ operator are:
\be
\varphi^{\psi',\psi}(x)=2\sum_{n=1}^{\infty}\left(\prod_{
		k=1\atop
		k\neq n}^{\infty}\delta_{r'_k,r_k}\right)\left(\sqrt{\frac{r_n}{n} }\,\delta_{r'_n+1,r_n} +\sqrt{\frac{r_n+1}{n}}\,\delta_{r'_n-1,r_n}\right)\sin(n{\pi}x/{L}).
\ee

\paragraph{Thermal states and time evolution -}

The density matrix $\rho(H,T)$ that describes a thermal state of some Hamiltonian $H$ at temperature $T$ is given by:
\be
\rho(H,T)=\frac{e^{-H/T}}{\text{tr}\left(e^{-H/T}\right)},
\ee
In order to construct a thermal density matrix, we first construct the TCSA form of $H$ and then compute $\rho(H,T)$ numerically, using matrix exponentiation.

Similarly for the study of quench dynamics, the time evolution operator:
\be
U(H,t)=e^{-\ii H t}.
\ee
corresponding to the chosen post-quench Hamiltonian $H$ and time $t$ after the quench, is computed numerically by matrix exponentiation.

\section{Truncation effects and computational performance} \label{sec:TruncationPerformance}

\paragraph{Truncation -}
The TCSA simulation is done by representing the operators as numerical matrices (using the matrix elements computed above) and introducing a cutoff. This is done by keeping only those states in the Hilbert space whose energy (with respect to the free boson Hamiltonian \eqref{eq:FBmatrixElements}) is below the chosen cutoff value. In this way we keep the matrices finite. The number of states in the Hilbert space for a chosen cutoff:
\be
\text{cutoff}:=\left(kr_k\right)_{\text{max}}
\ee
is given by the cumulative sum of the (combinatorial) partition function:
\be
\text{\#states}=\sum_{n=0}^{\text{cutoff}}p(n).
\ee
The values relevant for this work are listed in the table \ref{tab:NoStates}.

\begin{center}
	\begin{table}[htbp]
		\begin{tabular}{ |c|c| }
			\hline
			\textbf{cutoff} &  \,\textbf{\#states} \\ \hline
			15 & 684 \\ \hline
			16 & 915 \\ \hline
			17 & 1212 \\ \hline
			18 & 1597 \\ \hline
			19 & 2087 \\ \hline
			20 & 2714 \\ \hline
			21 & 3506 \\ \hline
			22 & 4508 \\ \hline
		\end{tabular}
		\caption{Number of states in the Hilbert space for the energy cutoff values used for the analysis in this work.\label{tab:NoStates}}
	\end{table}	
\end{center}

\paragraph{Truncation effects -}

Truncation effects originate from neglecting the contribution of modes above the cutoff energy. This means, on the one hand, that the highest spatial and temporal resolution we can achieve is restricted by the value of the energy cutoff, which plays also the role of a short-wavelength cutoff. On the other hand, since quantum dynamics is oscillatory and we have approximate values for the energy eigenvalues (therefore the oscillation frequencies), our time-series will eventually get out-of-phase after several oscillations, which restricts the longest time scale we can reach at a given cutoff. As in all spectral numerical methods, convergence for time-averaged values, amplitudes of oscillations and Fourier spectra is achieved much easier than for time-series data.

It should also be stressed that analogous effects are inevitably present in the experimental system \cite{exp-sG,exp-GGE}, since the SGM is only a low-energy approximation of the actual system. In such quantum gases the cutoff scale above which the approximation breaks down is determined by factors like the nonzero range of the effective inter-particle interaction, which induces nonlinearity in the dispersion relation of the bosonisation (density and phase) fields and violation of relativistic invariance at higher-energies. Therefore the challenge in comparing theory and experiment is precisely to disentangle such high-energy deviations from the low-energy physics.

There are several ways to determine the quality of our numerical data and the parameter ranges where the method is reliable. Expanding the states used in our computations in the free boson basis, one can examine the amplitudes versus the energies of the basis states and check whether they decrease to a sufficiently small value for states in the vicinity of the cutoff. Another way to verify the results is to plot the values of the observables (for example correlation function time-series) for different values of the cutoff and  check that they have converged within a sufficiently low tolerance level. In our computation we used both approaches to verify that the TCSA is reliable for the observables and parameter values used in this work.

\begin{figure}[htbp]
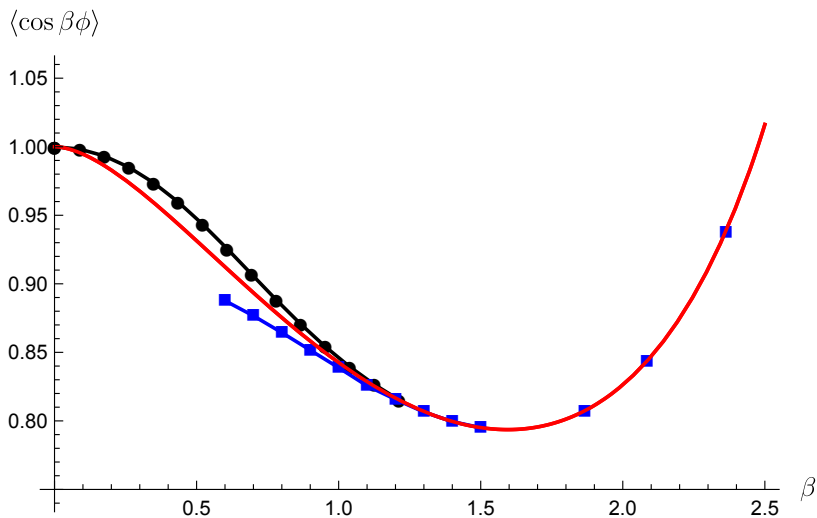

\begin{center}
 \centering
\includegraphics[width=.6\linewidth]{{{comparison}}}
\caption{Comparison of $\langle\cos\beta\phi\rangle$ as a function of interaction $\beta$ for three different types of states, computed by different methods: ground state in thermodynamically large system (red line) as given by the exact analytical formula of Lukyanov-Zamolodchikov \cite{Lukyanov-Zamo1}, ground state in a finite system of length $L=25$ with Dirichlet boundary conditions (black line and dots) computed from TCSA, ground state in a finite system of the same length with periodic boundary conditions (blue line and dots) computed numerically using the Non-Linear Integral Equation (NLIE) \cite{Klumper1991,Destri-deVega1}. The TCSA method gives reliable data for $\beta\lesssim1.2$, while the NLIE method for $\beta\gtrsim 0.6$. The three lines converge for $\beta \gtrsim 1$ ($\Delta \gtrsim 0.04$), because for such interactions the mass of the lightest breather is sufficiently larger than (at least 3 times) the inverse system size, so that the system is practically in the thermodynamic limit (finite size effects and dependence on boundary conditions is negligible). This convergence provides a nontrivial verification for our numerics.}
\label{fig:cos-beta-phi}
\end{center}
\end{figure}

In addition to the above convergence tests, we have also performed a number of nontrivial tests of our numerics through comparison with known analytical results and with other numerical methods. First, we compared the TCSA energy spectrum with that predicted by integrability following the approach in \cite{TCSA-bsG}. Second, we compared the expectation value of the vertex one-point function $\langle\cos\beta\phi\rangle$ to integrability predictions. We computed this value at the middle of the box with Dirichlet boundary conditions for various values of the interaction $\beta$ using the TCSA. For the ground state in an infinite size system, an analytical formula by Lukyanov-Zamolodchikov \cite{Lukyanov-Zamo1} is available. For the ground state value in a finite box with periodic boundary conditions one can use numerical data from the so-called Non-Linear Integral Equation \cite{Klumper1991,Destri-deVega1}. In Fig.~\ref{fig:cos-beta-phi} we plot together the results for these three different types of states as functions of the interaction. TCSA data converge well for all values of interaction $\beta\lesssim1.2$ ($\Delta\lesssim0.055$), while the NLIE converges for $\beta\gtrsim 0.6$. In the region $\beta \gtrsim 1$ ($\Delta \gtrsim 0.04$) where the correlation length is small enough compared to the system size so that finite size and boundary effects are negligible, all three methods give results that agree with each other very well. 
In order to benchmark the application of our numerical method to quench dynamics, we have also performed comparison with exact analytical results for the free massless and massive cases \cite{CC,CC2007} always observing good agreement. Analogous tests have been already performed successfully in the context of quenches in Ising field theory \cite{TCSA-QQ, TCSA_latest} and other truncation-based methods applied to the study of quantum quenches \cite{TCSA_QQ1,TCSA_QQ2,TCSA_QQ3,TCSA_QQ4}.

\paragraph{Performance -}
The crucial steps of our TCSA simulation are the computation of the matrix elements of the Hamiltonians and the observables (for the selected ordering of the states \eqref{eq:States} in the truncated Hilbert space), the diagonalization of the Hamiltonian to find the state of interest (or exponentiation in case of thermal states), the computation of the propagator over the chosen time step, the propagation of the state and the computation of expectation values of products of the observables (the correlators). All the steps apart from the last one are numerically cheap, since we only do them once (or once per time step in case of the propagation of the state). In particular, the matrices corresponding to the Hamiltonians and the observables can be computed once, saved to the hard drive and loaded when needed. The numerically most expensive step of the simulation is the computation of the correlation functions since we have to perform it (in each time step) as many times as the number of points in the grid with the chosen spatial resolution. For example, in our case, for the time evolution after a quench in each time step this amounts to $(2+2)\cdot41\times41\sim 6700$ matrix products for the 2-p correlators and $(2+4)\cdot41\times41\sim 10^4$ matrix products for the 4-p function, where the matrices are of the sizes given in table \ref{tab:NoStates}. For the cutoff of 20 the computation of a quench normally takes between a couple of days and a week on our computational cluster. The computation is much faster if one needs the time series at just a chosen point and does not have to evolve the full grid.

The most expensive computation in this work is the computation of the kurtosis, where to compute the 4-p functions over the full 4D grid, we need to perform $(2+4)\cdot21\times21\times21\times21\sim 10^6$ matrix products for each temperature and interaction. For this reason, the computation of the lines on Fig.~1 of the main text, take between a couple of weeks and a month. For the computation of the quench timeseries of the kurtosis, using the full 4D grid at all time steps ($\sim 600$) to perform the numerical integration would be extremely expensive, so we used instead a random sampling of $10^3$ spatial points. The accuracy of this method was verified by comparison to the full 4D result at the initial and a couple of random times. 
	
The memory usage is never an issue in our case, since because of performing so many matrix products, we are limited to the use of matrices of sufficiently small sizes that allow these operations to be performed fast enough. So the main resource needed is the processor power. One could further optimize the performance of the algorithm by taking into account the symmetries of the correlation functions. 


\section{Correlation functions}

As explained in the main text, multipoint correlation functions provide important physical information for a quantum field theory. In this work we are computing two- and four-point ($N=2,4$) equal-time correlation functions:
\be
G^{(N)}(x_1,x_2,\ldots,x_N)=\left\langle \varphi(x_1)\varphi(x_2)\cdots\varphi(x_N)\right\rangle
\ee
where the expectation value is taken either in an equilibrium state (of some Hamiltonian $H$ under consideration) or in time evolved states after a quench. Equilibrium states are either pure states $|\Psi\rangle$ (ground or excited states of $H$), in which case the expectation values are $\langle \dots \rangle = \langle \Psi|\dots |\Psi\rangle$ or mixed states $\rho$, like the thermal states we consider here, in which case $\langle \dots \rangle = \text{tr}\left( \dots \rho\right)$. In equilibrium states of the SGM all the odd order (odd $N$) correlation functions vanish, since the field \eqref{eq:FBfield} is odd under reflection ($\varphi(x)=-\varphi(-x)$) and the SGM Hamiltonian \eqref{eq:SGactionZamolodchikov} is even ($H(\varphi)=H(-\varphi)$). 

For the study of dynamics after a quench, the expectation value refers to the time evolved state:
\be
|\Psi(t)\rangle = U(H,t) |\Psi \rangle = e^{-\ii H t} |\Psi \rangle
\ee
where $|\Psi \rangle$ is the quench initial state, that is an equilibrium (ground or excited) state of some Hamiltonian $H_0$ (the pre-quench Hamiltonian),  and the time evolution operator $U(H,t)=e^{-\ii H t}$ corresponds to a different Hamiltonian $H$ (the post-quench Hamiltonian). In the dynamical case, we often denote the time dependent correlation functions as $G^{(N)}(x_1,x_2,\ldots,x_N;t)$. For the quenches considered here, the pre-quench Hamiltonian is the SGM Hamiltonian at some value of the interaction $\Delta_0$ and we study initial states that are either excited states or the ground state. The post-quench Hamiltonian, on the other hand, is the SGM corresponding to a different value of the interaction $\Delta$. For comparison, we also compute the quench dynamics of the same initial state under the free massless or massive Hamiltonian.

\paragraph{Connected correlation functions - }

If we are interested in studying only the genuine multiparticle interactions, we have to subtract from the full correlation function $G^{(N)}$ the contributions that come from lower order correlation functions (fewer particle collisions). One gets what is called the connected part of the correlation functions, which are essentially the joint cumulants of the fields in the state under consideration
\be
G_{\text{con}}^{(N)}(x_1,x_2,\ldots,x_N)=\sum_{\pi}\left[\left(|\pi|-1\right)!(-1)^{|\pi|-1}\prod_{B\in\pi}\left\langle\prod_{i\in B}\varphi(x_i)\right\rangle\right].
\ee
Here, $\pi$ are all possible partitions of $\{1,2,\ldots,N\}$ into blocks $B$ and $i$ are elements of $B$. $|\pi|$ is the number of blocks in the partition. All the correlation functions can be taken at time $t$. If all connected correlations of order higher than two vanish, then Wick's theorem holds and the system is free (i.e. noninteracting).

In case of four-point functions and vanishing odd correlation functions, this formula simplifies to:
\begin{eqnarray}
G_{\text{con}}^{(4)}(x_1,x_2,x_3,x_4)&=&G^{(4)}(x_1,x_2,x_3,x_4)-\nonumber\\
&&-G^{(2)}(x_1,x_2)G^{(2)}(x_3,x_4)-G^{(2)}(x_1,x_3)G^{(2)}(x_2,x_4)-G^{(2)}(x_1,x_4)G^{(2)}(x_2,x_3).
\end{eqnarray}

\paragraph{Kurtosis - }

To estimate how close the states are to Gaussian, that is, to see the strength of interaction effects, we compute the kurtosis -- the ratio between the integrated connected and full four-point correlation function \cite{exp-sG}:
\be
\mathcal{K}:=\frac{\int \dd x_1 \dd x_2 \dd x_3 \dd x_4 \left|G_{\text{con}}^{(4)}(x_1,x_2,x_3,x_4)\right|}{\int \dd x_1 \dd x_2 \dd x_3 \dd x_4 \left|G^{(4)}(x_1,x_2,x_3,x_4)\right|}\approx\frac{\sum_{x_1,x_2,x_3,x_4} \left|G_{\text{con}}^{(4)}(x_1,x_2,x_3,x_4)\right|}{\sum_{x_1,x_2,x_3,x_4} \left|G^{(4)}(x_1,x_2,x_3,x_4)\right|},
\ee
where in the last equality we used that in the numerical simulation the domain is discretised so the integral is approximated with a sum. For Gaussian states, $\mathcal{K}$ vanishes, while a larger value of $\mathcal{K}$ corresponds to a more strongly interacting system.

For the evaluation of the kurtosis one needs the 4-p correlated functions over the entire four dimensional grid of spatial positions. However, as explained in Section  "Truncation effects and computational performance", this is computationally very expensive, so for quench time sequences (i.e. time evolved states) we used a random sampling method over a thousand points, checking at a few time slices that it correctly reproduces the kurtosis result obtained from the full grid. For the value of the kurtosis in the initial state of the quench shown in Fig. 5 of the main text and for the thermal state plot (Fig. 1 in the main text) we used the full 4D grid for the computation.

\section{Quench from a ground state}

\begin{figure}[htbp]
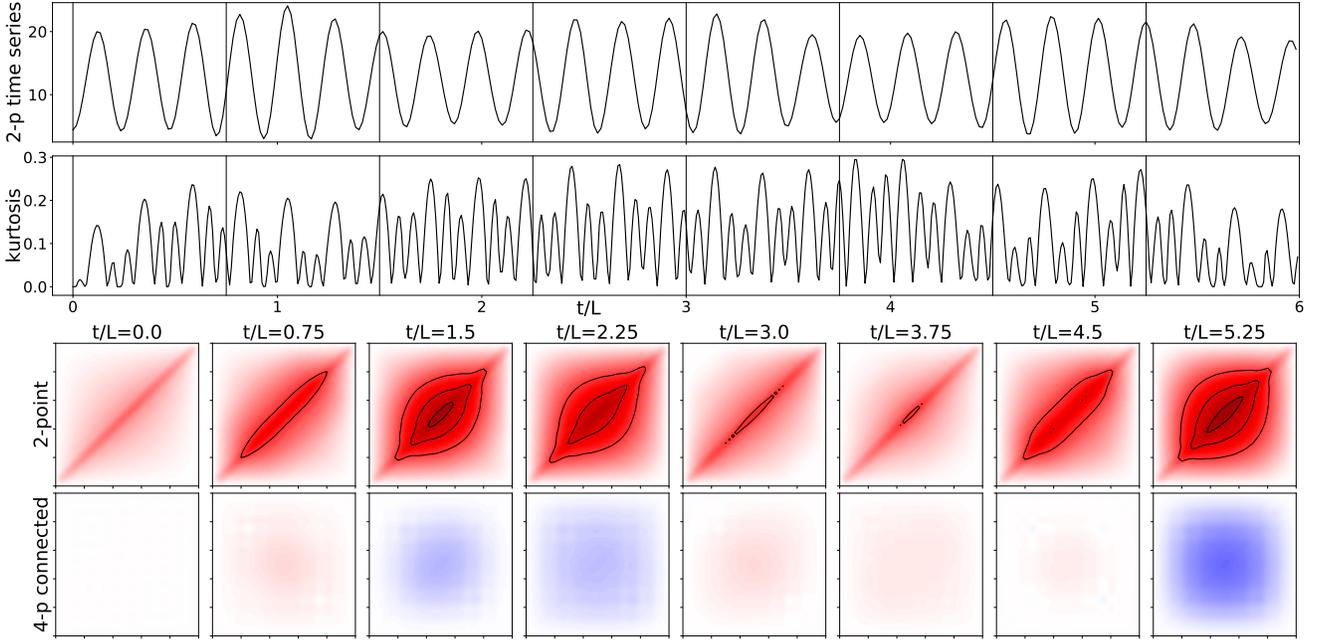

\begin{center}
 \centering
\includegraphics[width=.97\textwidth]{{{Plot_FigurePaper_SGDtoSGDquench_Coupling_Phi_labmda0_17_lambda_7_l_25_nPos_41_nHam_1_tM0_0_dtM_0.5_tMmax500_KRK_19}}}
\caption{{Time evolution of 2-p correlations $G^{(2)}(L/3,2L/3;t)$ and the kurtosis (\emph{top two rows}) and spatial density plots of 2-p correlations $G^{(2)}(x_1,x_2;t)$ and 4-p connected correlations $G^{(4)}_{\text{con}}(x_1,x_2,L/4,3L/4;t)$ at various times $t$ (\emph{bottom two rows}) after an interaction quench starting from a ground state of the SGM (pre-quench interaction:  $\Delta_0 =1/{18}$, post-quench interaction: $\Delta =1/{8}$, $L=25/M$).}}
\label{fig:quench2}
\end{center}
\end{figure}

In this section we present an interaction quench starting from the ground state of the SGM for $\Delta_0=1/18$ and quenching to 
$\Delta=1/8$ which is shown in Fig.~\ref{fig:quench2}.  
For comparison, the energy of this initial state is $\sim 0.111 \, M$ above the post-quench ground state, while the energy of the excited state shown in Fig.~5 of the main text is $\sim 1.243 \, M$, that is about 10 times higher. 
In contrast to the case of quench starting from an excited initial state, studied in the main text, in the present case the quench dynamics is dominated by low energy modes, the leading one being the lowest lying second breather mode (due to parity invariance the odd states do not contribute).

\begin{figure}[htbp]
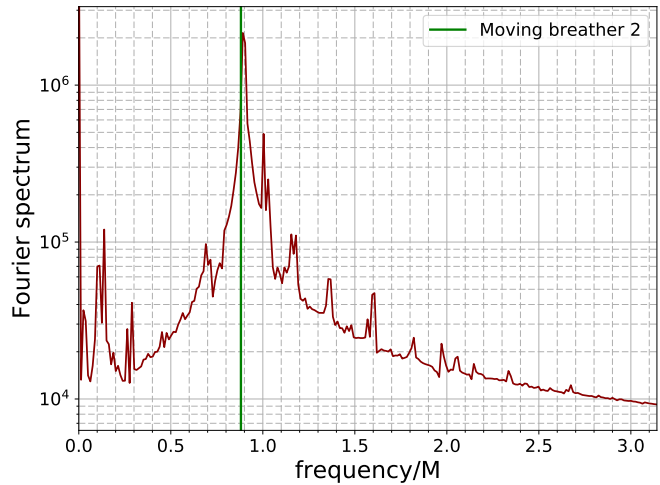

\begin{center}
 \centering
\includegraphics[width=.5\textwidth]{{{Plot_SGDtoSGDquench_Coupling_Phi_TwoP_labmda0_17_lambda_7_l_25_nPos_41_nHam_1_tM0_0_dtM_0.5_tMmax500_KRK_19}}}
\caption{Fourier spectrum of the time dependence of the spatially integrated 2-p correlations after the ground state quench shown in Fig.~\ref{fig:quench2}.}
\label{fig:spectrum}
\end{center}
\end{figure}

The Fourier spectrum of the time evolution of observables is determined by the post-quench excitations, which are multiparticle states with momenta quantised by the finite system size $L$. The complete set of equations that determine the energy levels can be found e.g. in \cite{TCSA-bsG}. Here we focus only on the dominant energy level which corresponds to the $n=2$ breather moving with the lowest momentum $p(\theta):=m_2\sinh\theta$ allowed by the Bethe Yang equations:
\be
e^{2ip(\theta)L} R(\theta)^2=+1
\ee
where the second breather reflection factor for Dirichlet boundary conditions is \cite{Ghoshal-Zamo,Ghoshal}: 
\begin{equation}
R(\theta)= \frac{\big(\frac12\big)_\theta\big(1+\frac\xi2\big)_\theta\big(\frac\xi2\big)_\theta\big(1+\xi\big)_\theta}{\big(\frac12+\frac\xi2\big)_\theta^2\big(\frac12-\frac\xi2\big)_\theta^2 \big(\frac32+\xi\big)_\theta\big(\frac32+\frac\xi2\big)_\theta^2}
\end{equation}
and the notation
\be
\big(x\big)_\theta := \frac{\sinh\left(\frac\theta 2+\frac{\ii \pi x}2\right)}{\sinh\left(\frac\theta 2-\frac{\ii \pi x}2\right)}
\ee
has been used. Note that static breathers are not present for Dirichlet boundary conditions. From these equations 
we find that the energy of this mode measured from the ground state is $\sim 0.881039 \, M$, which matches with the frequency of the dominant peak in the Fourier spectrum of 2-p correlations, shown in Fig.~\ref{fig:spectrum}.



\begin{thebibliography}{10}

\bibitem{Weinberg}
S.~Weinberg, \href{http://dx.doi.org/10.1017/cbo9781139644167}{{\em The Quantum
  Theory of Fields}}.
\newblock Cambridge University Press, 1995.

\bibitem{SM}
{See Supplementary Material for details.}

\bibitem{exp-sG}
T.~Schweigler, V.~Kasper, S.~Erne, I.~Mazets, B.~Rauer, F.~Cataldini,
  T.~Langen, T.~Gasenzer, J.~Berges, and J.~Schmiedmayer, ``Experimental
  characterization of a quantum many-body system via higher-order
  correlations,'' \href{http://dx.doi.org/10.1038/nature22310}{{\em Nature}
  {\bfseries 545} no.~7654, (2017) 323--326}.

\bibitem{exp-GGE}
T.~Langen, S.~Erne, R.~Geiger, B.~Rauer, T.~Schweigler, M.~Kuhnert,
  W.~Rohringer, I.~E. Mazets, T.~Gasenzer, and J.~Schmiedmayer, ``{Experimental
  observation of a generalized Gibbs ensemble},''
  \href{http://dx.doi.org/10.1126/science.1257026}{{\em Science} {\bfseries
  348} no.~6231, (2015) 207--211}.

\bibitem{1d-phys-rev}
M.~A. Cazalilla, R.~Citro, T.~Giamarchi, E.~Orignac, and M.~Rigol, ``One
  dimensional bosons: From condensed matter systems to ultracold gases,''
  \href{http://dx.doi.org/10.1103/revmodphys.83.1405}{{\em Reviews of Modern
  Physics} {\bfseries 83} no.~4, (2011) 1405--1466}.

\bibitem{Haldane}
F.~D.~M. Haldane, ``{`Luttinger liquid theory' of one-dimensional quantum
  fluids. I. Properties of the Luttinger model and their extension to the
  general 1D interacting spinless Fermi gas},''
  \href{http://dx.doi.org/10.1088/0022-3719/14/19/010}{{\em Journal of Physics
  C: Solid State Physics} {\bfseries 14} no.~19, (1981) 2585--2609}.

\bibitem{Luttinger-liquids}
V.~Mastropietro and D.~C. Mattis, {\em {Luttinger Model: The First 50 Years and
  Some New Directions (Series on Directions in Condensed Matter Physics)}}.
\newblock World Scientific Publishing Company, 2013.

\bibitem{exp:Josephson}
F.~S. Cataliotti, S.~Burger, C.~Fort, P.~Maddaloni, F.~Minardi, A.~Trombettoni,
  A.~Smerzi, and M.~Inguscio, ``{Josephson Junction Arrays with Bose-Einstein
  Condensates},'' \href{http://dx.doi.org/10.1126/science.1062612}{{\em
  Science} {\bfseries 293} no.~5531, (2001) 843--846}.

\bibitem{Gritsev2006}
V.~Gritsev, E.~Altman, E.~Demler, and A.~Polkovnikov, ``{Full quantum
  distribution of contrast in interference experiments between interacting
  one-dimensional Bose liquids},''
  \href{http://dx.doi.org/10.1038/nphys410}{{\em Nature Physics} {\bfseries 2}
  no.~10, (2006) 705--709}.

\bibitem{Gritsev2007a}
V.~Gritsev, E.~Demler, M.~Lukin, and A.~Polkovnikov, ``{Spectroscopy of
  Collective Excitations in Interacting Low-Dimensional Many-Body Systems Using
  Quench Dynamics},''
  \href{http://dx.doi.org/10.1103/physrevlett.99.200404}{{\em Physical Review
  Letters} {\bfseries 99} no.~20, (2007) 200404}.

\bibitem{Gritsev2007b}
V.~Gritsev, A.~Polkovnikov, and E.~Demler, ``Linear response theory for a pair
  of coupled one-dimensional condensates of interacting atoms,''
  \href{http://dx.doi.org/10.1103/physrevb.75.174511}{{\em Physical Review B}
  {\bfseries 75} no.~17, (2007) 174511}.

\bibitem{new-exp}
M.~Pigneur, T.~Berrada, M.~Bonneau, T.~Schumm, E.~Demler, and J.~Schmiedmayer,
  ``Relaxation to a phase-locked equilibrium state in a one-dimensional bosonic
  Josephson junction,''
  \href{http://dx.doi.org/10.1103/physrevlett.120.173601}{{\em Physical Review
  Letters} {\bfseries 120} no.~17, (Apr, 2018) 173601}.

\bibitem{Amit}
D.~J. Amit, Y.~Y. Goldschmidt, and S.~Grinstein, ``{Renormalisation group
  analysis of the phase transition in the 2D Coulomb gas, Sine-Gordon theory
  and {XY}-model},'' \href{http://dx.doi.org/10.1088/0305-4470/13/2/024}{{\em
  Journal of Physics A: Mathematical and General} {\bfseries 13} no.~2, (1980)
  585--620}.

\bibitem{Faddeev-Korepin_1}
V.~E. Korepin and L.~D. Faddeev, ``Quantization of solitons,''
  \href{http://dx.doi.org/10.1007/bf01028946}{{\em Theoretical and Mathematical
  Physics} {\bfseries 25} no.~2, (1975) 1039--1049}.

\bibitem{Faddeev-Korepin_2}
L.~Faddeev and V.~Korepin, ``Quantum theory of solitons,''
  \href{http://dx.doi.org/10.1016/0370-1573(78)90058-3}{{\em Physics Reports}
  {\bfseries 42} no.~1, (1978) 1--87}.

\bibitem{Giamarchi}
T.~Giamarchi, {\em {Quantum Physics in One Dimension}}.
\newblock {Oxford University Press}, 2003.

\bibitem{Essler-Konik}
F.~H. Essler and R.~M. Konik,
  \href{http://dx.doi.org/10.1142/9789812775344_0020}{``{Applications of
  Massive Integrable Quantum Field Theories to Problems in Condensed Matter
  Physics},''} in {\em From Fields to Strings: Circumnavigating Theoretical
  Physics}, pp.~684--830.
\newblock {World Scientific}, 2005.

\bibitem{Yurov-Zamo}
V.~P. Yurov and A.~B. Zamolodchikov, ``{Truncated Conformal Space Approach to
  scaling Lee-Yang model},''
  \href{http://dx.doi.org/10.1142/s0217751x9000218x}{{\em International Journal
  of Modern Physics A} {\bfseries 05} no.~16, (1990) 3221--3245}.

\bibitem{TCSA-3cIsing}
M.~L\"{a}ssig, G.~Mussardo, and J.~L. Cardy, ``{The scaling region of the
  tricritical Ising model in two dimensions},''
  \href{http://dx.doi.org/10.1016/0550-3213(91)90206-d}{{\em Nuclear Physics B}
  {\bfseries 348} no.~3, (1991) 591--618}.

\bibitem{TCSA-Smatrix2}
V.~Yurov and A.~Zamolodchikov, ``{Truncated-fermionic-space approach to the
  critical 2d Ising model with magnetic field},''
  \href{http://dx.doi.org/10.1142/s0217751x91002161}{{\em International Journal
  of Modern Physics A} {\bfseries 06} no.~25, (1991) 4557--4578}.

\bibitem{TCSA-sG1}
G.~Feverati, F.~Ravanini, and G.~Tak{\'{a}}cs, ``{Truncated conformal space at
  $c=1$, nonlinear integral equation and quantization rules for multi-soliton
  states},'' \href{http://dx.doi.org/10.1016/s0370-2693(98)00543-7}{{\em
  Physics Letters B} {\bfseries 430} no.~3-4, (1998) 264--273}.

\bibitem{TCSA-hd}
M.~Hogervorst, S.~Rychkov, and B.~C. van Rees, ``{Truncated conformal space
  approach in $d$ dimensions: A cheap alternative to lattice field theory?},''
  \href{http://dx.doi.org/10.1103/physrevd.91.025005}{{\em Physical Review D}
  {\bfseries 91} no.~2, (2015) 025005}.

\bibitem{TCSA-sG2}
G.~Feverati, F.~Ravanini, and G.~Tak{\'{a}}cs, ``{Non-linear integral equation
  and finite volume spectrum of sine-Gordon theory},''
  \href{http://dx.doi.org/10.1016/s0550-3213(98)00747-0}{{\em Nuclear Physics
  B} {\bfseries 540} no.~3, (1999) 543--586}.

\bibitem{TCSA-sG3}
G.~Feh{\'{e}}r and G.~Tak{\'{a}}cs, ``{Sine-Gordon form factors in finite
  volume},'' \href{http://dx.doi.org/10.1016/j.nuclphysb.2011.06.020}{{\em
  Nuclear Physics B} {\bfseries 852} no.~2, (2011) 441--467}.

\bibitem{Sachdev}
K.~Damle and S.~Sachdev, ``{Universal Relaxational Dynamics of Gapped
  One-Dimensional Models in the Quantum Sine-Gordon Universality Class},''
  \href{http://dx.doi.org/10.1103/physrevlett.95.187201}{{\em Physical Review
  Letters} {\bfseries 95} no.~18, (2005) 187201}.

\bibitem{Schweigler1}
S.~Beck, I.~E. Mazets, and T.~Schweigler, ``{A Non-Perturbative Method to
  compute Thermal Correlations in One-Dimensional Systems},''
  \href{http://arxiv.org/abs/1712.01190}{{\ttfamily arXiv:1712.01190}}.

\bibitem{Schweigler2}
S.~Beck, I.~E. Mazets, and T.~Schweigler, ``{Nonperturbative method to compute 
  thermal correlations in one-dimensional systems},''
  \href{http://dx.doi.org/10.1103/physreva.98.023613}{{\em Physical Review A} 
  {\bfseries 98} no.~2, (2018) 023613}.

\bibitem{Bertini-Essler}
B.~Bertini, D.~Schuricht, and F.~H.~L. Essler, ``{Quantum quench in the
  sine-Gordon model},''
  \href{http://dx.doi.org/10.1088/1742-5468/2014/10/p10035}{{\em Journal of
  Statistical Mechanics: Theory and Experiment} {\bfseries 2014} no.~10, (2014)
  P10035}.

\bibitem{Foini2015}
L.~Foini and T.~Giamarchi, ``{Nonequilibrium dynamics of coupled Luttinger
  liquids},'' \href{http://dx.doi.org/10.1103/physreva.91.023627}{{\em Physical
  Review A} {\bfseries 91} no.~2, (2015) 023627}.

\bibitem{Foini2017a}
L.~Foini and T.~Giamarchi, ``{Relaxation dynamics of two coherently coupled
  one-dimensional bosonic gases},''
  \href{http://dx.doi.org/10.1140/epjst/e2016-60383-x}{{\em The European
  Physical Journal Special Topics} {\bfseries 226} no.~12, (2017) 2763--2774}.

\bibitem{Foini2017b}
L.~Foini, A.~Gambassi, R.~Konik, and L.~F. Cugliandolo, ``{Measuring effective
  temperatures in a generalized Gibbs ensemble},''
  \href{http://dx.doi.org/10.1103/physreve.95.052116}{{\em Physical Review E}
  {\bfseries 95} no.~5, (2017) 052116}.

\bibitem{DeNardis2017}
J.~D. Nardis, M.~Panfil, A.~Gambassi, R.~Konik, L.~Cugliandolo, and L.~Foini,
  ``{Probing non-thermal density fluctuations in the one-dimensional Bose
  gas},'' \href{http://dx.doi.org/10.21468/scipostphys.3.3.023}{{\em {SciPost}
  Physics} {\bfseries 3} no.~3, (2017) 023}.

\bibitem{Kormos1}
M.~Kormos and G.~Zar{\'{a}}nd, ``{Quantum quenches in the sine-Gordon model: A
  semiclassical approach},''
  \href{http://dx.doi.org/10.1103/physreve.93.062101}{{\em Physical Review E}
  {\bfseries 93} no.~6, (2016) 062101}.

\bibitem{Kormos2}
C.~P. Moca, M.~Kormos, and G.~Zar{\'{a}}nd, ``{Hybrid Semiclassical Theory of
  Quantum Quenches in One-Dimensional Systems},''
  \href{http://dx.doi.org/10.1103/physrevlett.119.100603}{{\em Physical Review
  Letters} {\bfseries 119} no.~10, (2017) 100603}.

\bibitem{Zamo-Zamo}
A.~B. Zamolodchikov and A.~B. Zamolodchikov, ``{Factorized S-matrices in two
  dimensions as the exact solutions of certain relativistic quantum field
  theory models},'' \href{http://dx.doi.org/10.1016/0003-4916(79)90391-9}{{\em
  Annals of Physics} {\bfseries 120} no.~2, (1979) 253--291}.

\bibitem{DHN}
R.~F. Dashen, B.~Hasslacher, and A.~Neveu, ``Particle spectrum in model field
  theories from semiclassical functional integral techniques,''
  \href{http://dx.doi.org/10.1103/physrevd.11.3424}{{\em Physical Review D}
  {\bfseries 11} no.~12, (1975) 3424--3450}.

\bibitem{Zamo77}
A.~B. Zamolodchikov, ``{Exact two-particle {S}-matrix of quantum sine-Gordon
  solitons},'' \href{http://dx.doi.org/10.1007/bf01626520}{{\em Communications
  in Mathematical Physics} {\bfseries 55} no.~2, (1977) 183--186}.

\bibitem{Lukyanov-Zamo1}
S.~Lukyanov and A.~Zamolodchikov, ``{Exact expectation values of local fields
  in the quantum sine-Gordon model},''
  \href{http://dx.doi.org/10.1016/s0550-3213(97)00123-5}{{\em Nuclear Physics
  B} {\bfseries 493} no.~3, (1997) 571--587}.

\bibitem{Smirnov}
F.~A. Smirnov, {\em {Form Factors in Completely Integrable Models of Quantum
  Field Theory (Advanced Series in Mathematical Physics)}}.
\newblock World Scientific Pub Co Inc, 1992.

\bibitem{LeClair-Mussardo}
A.~LeClair and G.~Mussardo, ``Finite temperature correlation functions in
  integrable {QFT},''
  \href{http://dx.doi.org/10.1016/s0550-3213(99)00280-1}{{\em Nuclear Physics
  B} {\bfseries 552} no.~3, (1999) 624--642}.

\bibitem{Pozsgay2011}
B.~Pozsgay, ``{Mean values of local operators in highly excited Bethe
  states},'' \href{http://dx.doi.org/10.1088/1742-5468/2011/01/p01011}{{\em
  Journal of Statistical Mechanics: Theory and Experiment} {\bfseries 2011}
  no.~01, (2011) P01011}.

\bibitem{Buccheri}
F.~Buccheri and G.~Tak{\'{a}}cs, ``{Finite temperature one-point functions in
  non-diagonal integrable field theories: the sine-Gordon model},''
  \href{http://dx.doi.org/10.1007/jhep03(2014)026}{{\em Journal of High Energy
  Physics} {\bfseries 2014} no.~3, (2014) 026}.

\bibitem{Fioretto}
D.~Fioretto and G.~Mussardo, ``Quantum quenches in integrable field theories,''
  \href{http://dx.doi.org/10.1088/1367-2630/12/5/055015}{{\em New Journal of
  Physics} {\bfseries 12} no.~5, (2010) 055015}.

\bibitem{Cubero-Schuricht}
A.~C. Cubero and D.~Schuricht, ``{Quantum quench in the attractive regime of
  the sine-Gordon model},''
  \href{http://dx.doi.org/10.1088/1742-5468/aa8c2e}{{\em Journal of Statistical
  Mechanics: Theory and Experiment} {\bfseries 2017} no.~10, (2017) 103106}.

\bibitem{CFT}
P.~D. Francesco, P.~Mathieu, and D.~S{\'{e}}n{\'{e}}chal,
  \href{http://dx.doi.org/10.1007/978-1-4612-2256-9}{{\em Conformal Field
  Theory}}.
\newblock Springer New York, 1997.

\bibitem{Ghoshal-Zamo}
S.~Ghoshal and A.~Zamolodchikov, ``{Boundary S-Matrix and Boundary State in
  Two-Dimensional Integrable Quantum Field Theory},''
  \href{http://dx.doi.org/10.1142/s0217751x94001552}{{\em International Journal
  of Modern Physics A} {\bfseries 09} no.~21, (1994) 3841--3885}.

\bibitem{bsG1}
Z.~Bajnok, L.~Palla, and G.~Tak{\'{a}}cs, ``{Boundary states and finite size
  effects in sine-Gordon model with Neumann boundary condition},''
  \href{http://dx.doi.org/10.1016/s0550-3213(01)00391-1}{{\em Nuclear Physics
  B} {\bfseries 614} no.~3, (2001) 405--448}.

\bibitem{TCSA-bsG}
Z.~Bajnok, L.~Palla, and G.~Tak{\'{a}}cs, ``{Finite size effects in boundary
  sine-Gordon theory},''
  \href{http://dx.doi.org/10.1016/s0550-3213(01)00616-2}{{\em Nuclear Physics
  B} {\bfseries 622} no.~3, (2002) 565--592}.

\bibitem{Deutsch}
J.~M. Deutsch, ``Quantum statistical mechanics in a closed system,''
  \href{http://dx.doi.org/10.1103/physreva.43.2046}{{\em Physical Review A}
  {\bfseries 43} no.~4, (1991) 2046--2049}.

\bibitem{Srednicki}
M.~Srednicki, ``Chaos and quantum thermalization,''
  \href{http://dx.doi.org/10.1103/physreve.50.888}{{\em Physical Review E}
  {\bfseries 50} no.~2, (1994) 888--901}.

\bibitem{Rigol-Srednicki}
M.~Rigol and M.~Srednicki, ``{Alternatives to Eigenstate Thermalization},''
  \href{http://dx.doi.org/10.1103/physrevlett.108.110601}{{\em Physical Review
  Letters} {\bfseries 108} no.~11, (2012) 110601}.

\bibitem{vonNeumann1929}
J.~von Neumann, ``{Beweis des Ergodensatzes und des H-Theorems in der neuen
  Mechanik},'' \href{http://dx.doi.org/10.1007/bf01339852}{{\em Zeitschrift
  f\"{u}r Physik} {\bfseries 57} no.~1-2, (1929) 30--70}.

\bibitem{vonNeumann}
J.~von Neumann, ``{Proof of the ergodic theorem and the H-theorem in quantum
  mechanics},'' \href{http://dx.doi.org/10.1140/epjh/e2010-00008-5}{{\em The
  European Physical Journal H} {\bfseries 35} no.~2, (2010) 201--237}.

\bibitem{Gogolin-Eisert}
C.~Gogolin and J.~Eisert, ``Equilibration, thermalisation, and the emergence of
  statistical mechanics in closed quantum systems,''
  \href{http://dx.doi.org/10.1088/0034-4885/79/5/056001}{{\em Reports on
  Progress in Physics} {\bfseries 79} no.~5, (2016) 056001}.

\bibitem{Zamo1995}
A.~B. Zamolodchikov, ``{Mass scale in the sine-Gordon model and its
  reductions},'' \href{http://dx.doi.org/10.1142/s0217751x9500053x}{{\em
  International Journal of Modern Physics A} {\bfseries 10} no.~08, (1995)
  1125--1150}.

\bibitem{Klumper1991}
A.~Klumper, M.~T. Batchelor, and P.~A. Pearce, ``{Central charges of the 6- and
  19-vertex models with twisted boundary conditions},''
  \href{http://dx.doi.org/10.1088/0305-4470/24/13/025}{{\em Journal of Physics
  A: Mathematical and General} {\bfseries 24} no.~13, (1991) 3111--3133}.

\bibitem{Destri-deVega1}
C.~Destri and H.~J. de~Vega, ``{New thermodynamic Bethe ansatz equations
  without strings},'' \href{http://dx.doi.org/10.1103/physrevlett.69.2313}{{\em
  Physical Review Letters} {\bfseries 69} no.~16, (1992) 2313--2317}.

\bibitem{CC}
P.~Calabrese and J.~Cardy, ``{Time Dependence of Correlation Functions
  Following a Quantum Quench},''
  \href{http://dx.doi.org/10.1103/physrevlett.96.136801}{{\em Physical Review
  Letters} {\bfseries 96} no.~13, (Apr, 2006) 136801}.

\bibitem{CC2007}
P.~Calabrese and J.~Cardy, ``{Quantum quenches in extended systems},''
  \href{http://dx.doi.org/10.1088/1742-5468/2007/06/p06008}{{\em Journal of
  Statistical Mechanics: Theory and Experiment} {\bfseries 2007} no.~06, (Jun,
  2007) P06008}.

\bibitem{TCSA-QQ}
T.~Rakovszky, M.~Mesty{\'{a}}n, M.~Collura, M.~Kormos, and G.~Tak{\'{a}}cs,
  ``{Hamiltonian truncation approach to quenches in the Ising field theory},''
  \href{http://dx.doi.org/10.1016/j.nuclphysb.2016.08.024}{{\em Nuclear Physics
  B} {\bfseries 911} (Oct, 2016) 805--845}.

\bibitem{TCSA_latest}
K.~{H{\'o}ds{\'a}gi}, M.~{Kormos}, and G.~{Tak{\'a}cs}, ``{Quench dynamics of
  the Ising field theory in a magnetic field},''
  \href{http://arxiv.org/abs/1803.01158}{{\ttfamily arXiv:1803.01158}}.

\bibitem{TCSA_QQ1}
J.-S. Caux and R.~M. Konik, ``{Constructing the Generalized Gibbs Ensemble
  after a Quantum Quench},''
  \href{http://dx.doi.org/10.1103/physrevlett.109.175301}{{\em Physical Review
  Letters} {\bfseries 109} no.~17, (Oct, 2012) 175301}.

\bibitem{TCSA_QQ2}
A.~J.~A. James and R.~M. Konik, ``{Quantum quenches in two spatial dimensions
  using chain array matrix product states},''
  \href{http://dx.doi.org/10.1103/physrevb.92.161111}{{\em Physical Review B}
  {\bfseries 92} no.~16, (Oct, 2015) 161111}.

\bibitem{TCSA_QQ3}
G.~P. Brandino, J.-S. Caux, and R.~M. Konik, ``{Glimmers of a Quantum {KAM}
  Theorem: Insights from Quantum Quenches in One-Dimensional Bose Gases},''
  \href{http://dx.doi.org/10.1103/physrevx.5.041043}{{\em Physical Review X}
  {\bfseries 5} no.~4, (Dec, 2015) 041043}.

\bibitem{TCSA_QQ4}
N.~J. Robinson, J.-S. Caux, and R.~M. Konik, ``{Motion of a Distinguishable
  Impurity in the Bose Gas: Arrested Expansion Without a Lattice and Impurity
  Snaking},'' \href{http://dx.doi.org/10.1103/physrevlett.116.145302}{{\em
  Physical Review Letters} {\bfseries 116} no.~14, (Apr, 2016) 145302}.

\bibitem{Ghoshal}
S.~Ghoshal, ``{Bound State Boundary S-matrix of the sine-Gordon Model},''
  \href{http://dx.doi.org/10.1142/s0217751x94001941}{{\em International Journal
  of Modern Physics A} {\bfseries 09} no.~27, (1994) 4801--4810}.

\end{thebibliography}

\providecommand{\href}[2]{#2}

\end{document}